\documentstyle[12pt,a4]{article}
\begin{document}

\begin{center}
{\bf SUPERPARTNER STATES IN QUANTUM MECHANICS OF COLORED PARTICLE }
\end{center}

\begin{center}
Sh.Mamedov
\end{center}

\begin{center}
High Energy Physics Laboratory, Baku State University, \\
23 Z. Khalilov , 370148 Baku, Azerbaijan,
e-mail: sh\_mamedov@yahoo.com
\end{center}

\begin{abstract}
Superpartner correspodence of states of colored particle in external
chromomagnetic field given by non-commuting axial vector potentials is
determined. Squared Dirac equation for this particle is solved exactly and
explicit expressions of superpartner states are found. The wave functions of
states with definite energy are found. Supersymmetry algebra and
superpartner states in three dimensional motion case are considered.
\end{abstract}

Supersymmetry and superpartner states in abelian quantum mechanics is well
studied [1-3]. For example, electron moving in homogenous magnetic field has
superpartner states differing from each other by projection of spin and main
quantum number [3,11,12].\ 

Supersymmetry in quantum mechanics of colored particle in chromomagnetic
field given by non-commuting vector potentials was studied in paper [8]. It
was constructed supercharge operators $Q_{\pm }$ and was shown, that these
operators form closed algebra together with so-called Hamiltonian -the
square of Dirac operator in given field. Of course, there are superpartner
states - states with same energy and different quantum numbers of colored
particle in such external field. But energy spectrum in this case is
continous and bosonic operators of creation and annihilation are different
from that one's in abelian quantum mechanics. So, it has sense to find
superpartner states in this case as well. The supersymmetry in Dirac
equation shows up the spin diagonal form of Hamiltonian for this particle
[8]. This means the equations corresponding to different projections of
quark's spin are separeted and can be solved independently. In this way
Dirac equation can be solved and superpartner states could be found.

Let us define external chromomagnetic field by constant vector potentials
[5]. Within SU(3) color symmetry group they look as

\begin{equation}
\label{1}A_1^a=(0,\tau _1^{\frac 12},0,0),\ A_2^a=(0,0,\tau _2^{\frac
12},0),\ A_3^a=0,\ A_0^a=0 
\end{equation}
where $a=\overline{1,8}$ is a color index and $\tau _{1,2}$ is a constant.
The field (1) is directed along third axes of ordinary and color spaces :

\begin{equation}
\label{2}F_{12}^3=H_z^3=g\tau _1^{\frac 12}\tau _2^{\frac 12},\quad other\ \
F_{\mu \nu }^a=0. 
\end{equation}
Here $g$ is color interaction constant.

The Dirac equation for a colored particle in the external color field has a
form

\begin{equation}
\label{3}\left( \gamma ^\mu P_\mu -m\right) \psi =0, 
\end{equation}
where $P_\mu =p_\mu +gA_\mu =p_\mu +gA_\mu ^a\frac{\lambda ^a}2$ , $\lambda
^a$ is Gell-Mann matrices describing particle's color spin. The equation (3)
is divided into two equations for Maiorana spinors $\phi $ and $\chi $ , $%
\psi =\left( 
\begin{array}{c}
\phi \\ 
\chi 
\end{array}
\right) $

$$
\left( \sigma ^iP_i\right) \chi =\left( i\frac \partial {\partial
t}-m\right) \phi 
$$

\begin{equation}
\label{4}\left( \sigma ^iP_i\right) \phi =\left( i\frac \partial {\partial
t}+m\right) \chi , 
\end{equation}
where Pauli matrices $\sigma ^i$ describe particle's spin. The spinors $\phi 
$ and $\chi $ has two components corresponding to two spin states of a
particle

$$
\phi =\left( 
\begin{array}{c}
\phi (\sigma _3=+1) \\ 
\phi (\sigma _3=-1) 
\end{array}
\right) =\left( 
\begin{array}{c}
\phi _{+} \\ 
\phi _{\_} 
\end{array}
\right) ,\qquad \chi =\left( 
\begin{array}{c}
\chi (\sigma _3=+1) \\ 
\chi (\sigma _3=-1) 
\end{array}
\right) =\left( 
\begin{array}{c}
\chi _{+} \\ 
\chi _{-} 
\end{array}
\right) . 
$$
Each component of $\phi $ and $\chi $ transforms under fundamental
representation of color group SU(3) and has three color components
describing color states of a particle and corresponding to three eigenvalues
of color spin $\lambda ^3$

$$
\phi _{\pm }=\left( 
\begin{array}{c}
\phi _{\pm }(\lambda ^3=+1) \\ 
\phi _{\pm }(\lambda ^3=-1) \\ 
\phi _{\pm }(\lambda ^3=0) 
\end{array}
\right) =\left( 
\begin{array}{c}
\phi _{\pm }^{(1)} \\ 
\phi _{\pm }^{(2)} \\ 
\phi _{\pm }^{(3)} 
\end{array}
\right) ,\qquad \chi _{\pm }=\left( 
\begin{array}{c}
\chi _{\pm }(\lambda ^3=+1) \\ 
\chi _{\pm }(\lambda ^3=-1) \\ 
\chi _{\pm }(\lambda ^3=0) 
\end{array}
\right) =\left( 
\begin{array}{c}
\chi _{\pm }^{(1)} \\ 
\chi _{\pm }^{(2)} \\ 
\chi _{\pm }^{(3)} 
\end{array}
\right) . 
$$
Acting on equations (4) by the operator $\left( \sigma ^iP_i\right) $ and
taking them into account once more, we get the same equations for spinors $%
\phi $ and $\chi $

\begin{equation}
\label{5}H\psi =\left( \sigma ^iP_i\right) ^2\psi =-\left( \frac{\partial ^2 
}{\partial t^2}+m^2\right) \psi . 
\end{equation}
For the particle with $p_3=0$ the Hamiltonian $H$ in (5) and the supercahrge
operators $Q_{\pm }=P_{\mp }a_{\pm }$ form supersymmetry algebra $\left\{
Q_{+},Q_{-}\right\} =H$\footnote{for more detail see [8]}. Here the
operators of bosonic and fermionic states are defined by formulas

$$
P_{\pm }=P_1\pm iP_2,\ a_{\pm }=\frac 12\left( \sigma _1\pm i\sigma
_2\right) 
$$
and obey next relations

$$
\left[ P_{+},P_{-}\right] =\lambda ^3gH_z^3,\ \left\{ a_{+},a_{-}\right\}
=1. 
$$

As we know, the superpartner states are received in the result of
supercharge operators $Q_{\pm }$ action. The action of fermionic operators
of creation and annihilation is turning over ordinary spin of particle

\begin{equation}
\label{6} 
\begin{array}{c}
a_{+}\left( 
\begin{array}{c}
\psi _{+} \\ 
\psi _{-} 
\end{array}
\right) =\left( 
\begin{array}{cc}
0 & 1 \\ 
0 & 0 
\end{array}
\right) \left( 
\begin{array}{c}
\psi _{+} \\ 
\psi _{-} 
\end{array}
\right) =\left( 
\begin{array}{c}
\psi _{-} \\ 
0 
\end{array}
\right) \\ 
a_{-}\left( 
\begin{array}{c}
\psi _{+} \\ 
\psi _{-} 
\end{array}
\right) =\left( 
\begin{array}{cc}
0 & 0 \\ 
1 & 0 
\end{array}
\right) \left( 
\begin{array}{c}
\psi _{+} \\ 
\psi _{-} 
\end{array}
\right) =\left( 
\begin{array}{c}
0 \\ 
\psi _{+} 
\end{array}
\right) . 
\end{array}
\end{equation}
It is seen is from (6) that bosonic operator of creation $P_{+}$ act only at 
$\psi _{+}$ and annihilation operator $P_{-}$ only at $\psi _{-}$. These
operators mix color states $\psi ^{(1)}$ and $\psi ^{(2)}$

\begin{equation}
\label{7} 
\begin{array}{c}
P_{+}\left( 
\begin{array}{c}
\psi _{-}^{(1)} \\ 
\psi _{-}^{(2)} \\ 
\psi _{-}^{(3)} 
\end{array}
\right) =\left( 
\begin{array}{c}
\left( p_1+ip_2\right) \psi _{-}^{(1)}+\frac g2\left( \tau _1^{\frac
12}+\tau _2^{\frac 12}\right) \psi _{-}^{(2)} \\ 
\frac g2\left( \tau _1^{\frac 12}-\tau _2^{\frac 12}\right) \psi
_{-}^{(1)}+\left( p_1+ip_2\right) \psi _{-}^{(2)} \\ 
\left( p_1+ip_2\right) \psi _{-}^{(3)} 
\end{array}
\right) =\left( 
\begin{array}{c}
\Psi _{-}^{(1)} \\ 
\Psi _{-}^{(2)} \\ 
\Psi _{-}^{(3)} 
\end{array}
\right) \\ 
P_{\_}\left( 
\begin{array}{c}
\psi _{+}^{(1)} \\ 
\psi _{+}^{(2)} \\ 
\psi _{+}^{(3)} 
\end{array}
\right) =\left( 
\begin{array}{c}
\left( p_1-ip_2\right) \psi _{+}^{(1)}+\frac g2\left( \tau _1^{\frac
12}-\tau _2^{\frac 12}\right) \psi _{+}^{(2)} \\ 
\frac g2\left( \tau _1^{\frac 12}+\tau _2^{\frac 12}\right) \psi
_{+}^{(1)}+\left( p_1-ip_2\right) \psi _{+}^{(2)} \\ 
\left( p_1-ip_2\right) \psi _{+}^{(3)} 
\end{array}
\right) =\left( 
\begin{array}{c}
\Psi _{+}^{(1)} \\ 
\Psi _{+}^{(2)} \\ 
\Psi _{+}^{(3)} 
\end{array}
\right) . 
\end{array}
\end{equation}
In the result of this mixing the superpartner states $\Psi _{\pm }^{(1),(2)}$
of pure color states $\psi _{\pm }^{(1)}$ and $\psi _{\pm }^{(2)}$ will be
mixed color states.

By use of supercharge operators it is easy to observe the spin diagonal form
of Hamiltonian $H$

\begin{equation}
\label{8}H\left( 
\begin{array}{c}
\psi _{+} \\ 
\psi _{-} 
\end{array}
\right) =\left( 
\begin{array}{c}
P_{-}P_{+}\qquad 0 \\ 
0\qquad P_{+}P_{-} 
\end{array}
\right) \left( 
\begin{array}{c}
\psi _{+} \\ 
\psi _{-} 
\end{array}
\right) . 
\end{equation}
So, due to spin diagonal form of $H$ the equation (5) is splitting into two
non-mixed spin indicies + and - equations for components $\psi _{+}$ and $%
\psi _{-}$

$$
P_{-}P_{+}\psi _{+}=-\left( \frac{\partial ^2}{\partial t^2}+m^2\right) \psi
_{+} 
$$

\begin{equation}
\label{9}P_{+}P_{-}\psi _{-}=-\left( \frac{\partial ^2}{\partial t^2}%
+m^2\right) \psi _{-}. 
\end{equation}
Taking into account the fact that the states $\psi _{\pm }$ are stationary (
the field (1) is time independent ) $\frac{\partial \psi _{\pm }}{\partial t}%
=iE\psi _{\pm }$ , the explicit form of equations (9) will be

\begin{equation}
\label{10}\left[ p_1^2+p_2^2+\frac{g^2}4\left( \tau _1+\tau _2\right)
I_2+g\left( p_1\tau _1^{\frac 12}\lambda ^1+p_2\tau _2^{\frac 12}\lambda
^2\mp \frac{H_z^3}2\lambda ^3\right) \right] \psi _{\pm }=\left(
E^2-m^2\right) \psi _{\pm } 
\end{equation}

$\left( {\rm here \; }I_2=\left( 
\begin{array}{ccc}
1 & 0 & 0 \\ 
0 & 1 & 0 \\ 
0 & 0 & 0 
\end{array}
\right) { {\rm is\; color\; matrix} }\right) .$

The energy spectrum $E$ in the field (1) was found in [3] and equal to ( on $%
p_z=0$ )

\begin{equation}
\label{11} 
\begin{array}{c}
E_{1,2}^2=p_{\bot }^2+m^2+ 
\frac{g^2}4\left( \tau _1+\tau _2\right) \pm g\sqrt{\tau _1p_1^2+\tau _2p^2+ 
\frac{\left( H_z^3\right) ^2}4} \\ E_3^2=p_{\bot }^2+m^2,\ \ \left( p_{\bot
}^2=p_1^2+p_2^2\right) . 
\end{array}
\end{equation}
\qquad The energy spectrum $E_3$ corresponds to colorless particle state $%
\left( \lambda ^3=0\right) .$ So, the degeneration of $E_3$ is a spin
degeneration. The spectra $E_{1,2}^2$ are received from (10) solving it for $%
E$ in momentum representation. The sign at last term in the expression of $%
E_{1,2}^2$ is defined by the operator

\begin{equation}
\label{12}\left( I_{\pm }^a\lambda ^a\right) =g\left( p_1\tau _1^{\frac
12}\lambda ^1+p_2\tau ^{\frac 12}\lambda ^2\mp \frac{H_z^3}2\lambda
^3\right) 
\end{equation}
square of which is same for spin\ $+$ and $-$ indices

\begin{equation}
\label{13}\left( I_{\pm }^a\lambda ^a\right) ^2=g^2\left( p_1^2\tau
_1+p_2^2\tau _2+\frac{\left( H_z^3\right) ^2}4\right) 
\end{equation}
and has eigenvalues $\pm \sqrt{p_1^2\tau _1+p_2^2\tau _2+\frac{\left(
H_z^3\right) ^2}4}$%
\footnote{Sign before the root shouldn't be confused with the spin + and - indices.}%
. The operator $\left( I_{\pm }^a\lambda ^a\right) $ is different for $\psi
_{+{ }}$ and $\psi _{-},$ and the superpartner Hamiltonians in (10) are
different for them as well. But solving characteristic equation gives same
energy spectrum branchs (11) for both of them. So, both energy branchs $%
E_{1,2{ }}$ are degenerated by ordinary spin [7]. It becomes clear that
energy spectrum $E_{1,2{ }}$is not defined neither by spin nor by color
quantum numbers 
\footnote{The quantum nimber  $\lambda ^3$ is not a conserved  quantity in the field (1) too $\left[ H,\lambda ^3\right] \neq 0$ .}%
. Energy branchs $E_{1{ }}$and $E_{2{ }}$ are defined only by
eigenvalues of $\left( I_{\pm }^a\lambda ^a\right) $ operator. So, both
energy branch are fourfold degenerated. This means that both spin and color
states $\psi _{\pm }^{(1)}$ and $\psi _{\pm }^{(2)}$ may be in the energy
branch $E_1$ or $E_{2{ }}$and the equations (10) for given energy $E_{1 
{ }}$ or $E_{2{ }}$should be solved for all spin and color states.
The operator $\left( I_{\pm }^a\lambda ^a\right) $ mixes the color states $%
\psi ^{(1)}$ and $\psi ^{(2)}$. But the solutions of equation (10) could be
received for pure color states because of color diagonal form of (13).

In order to solve equations (10) it is reasonable to use polar coordinates $%
x=r\cos \theta ,\ y=r\sin \theta $. We have two variables $\left( r,\theta
\right) $ and one constraint due to the conservation of quantity $\left(
\sigma ^iP_i\right) $ $\quad \left( \ \left[ H,\left( \sigma ^iP_i\right)
\right] =0,\quad {see [8] as well\ }\right) .$ In reference frame,
where $\theta $ is angle between $\overrightarrow{\sigma }$ and $
\overrightarrow{P},$ $\cos \theta =\frac{\left( \overrightarrow{\sigma }, 
\overrightarrow{P}\right) }{\mid \overrightarrow{\sigma }\mid \mid 
\overrightarrow{P}\mid }$ is a constant. Consequently $\theta $ is constant
too and $\frac{\partial \psi _{\pm }^{(1),(2)}}{\partial \theta }=0.$ Then
equations for $\psi _{+}$ look as

\begin{equation}
\label{14}\left\{ 
\begin{array}{c}
\left( 
\frac{d^2}{dr^2}+A_1\right) \psi _{+}^{(1)}+B_1\frac{d\psi _{+}^{(2)}}{dr}=0
\\ \left( 
\frac{d^2}{dr^2}+A_2\right) \psi _{+}^{(2)}+B_2\frac{d\psi _{+}^{(1)}}{dr}=0
\\ \left( \frac{d^2}{dr^2}+\left( E_3^2-m^2\right) \right) \psi _{+}^{(3)}=0 
\end{array}
\right. 
\end{equation}

$$
A_{1,2}=E^2-m^2-\frac{g^2}4\left( \tau _1^{\frac 12}\mp \tau _2^{\frac
12}\right) ^2,\ B_{1,2}=ig\left( \tau _1^{\frac 12}\cos \theta \ \mp i\tau
_2^{\frac 12}\sin \theta \right) 
$$
from which we get equation for a pure state $\psi _{+}^{(2)}$

\begin{equation}
\label{15}\frac{d^4}{dr^4}\psi _{+}^{(2)}+\left( A_1+A_2-B_1B_2\right) \frac{%
d^2}{dr^2}\psi _{+}^{(2)}+A_1A_2\psi _{+}^{(2)}=0. 
\end{equation}
The solution of (15) will be in the form $\psi _{+}^{(2)}(r)\sim e^{\alpha
r}.$ Putting this solution in equation (15) we find next expressions for $%
\alpha $ 
$$
\begin{array}{c}
\alpha _{1,2}=\pm \ ip_{\bot }, \\ 
\alpha _{3,4}=\pm i\sqrt{p_{\bot }^2+(\pm )2g\sqrt{p_{\bot }^2(\tau _1\cos
{}^2\theta +\tau _2\sin {}^2\theta )+\frac{(H_z^3)^2}4\ }+g^2(\tau _1\cos
{}^2\theta +\tau _2\sin {}^2\theta )\ }. 
\end{array}
$$
The sign in the bracket is the sign in the energy spectrum. The solution $%
\psi _{+}^{(2)}$ has a form

\begin{equation}
\label{16}\psi _{+}^{(2)}=\sum C_ie^{\alpha _ir}. 
\end{equation}
Here $C_i\,\,$is\thinspace arbitrary constants.\thinspace The wave function $%
\psi _{+}^{(1)}$ can be found from (12) using solution $\psi _{+}^{(2)}$ as
well

\begin{equation}
\label{17}\psi _{+}^{(1)}=-\frac 1{B_2}\sum C_i\left( \alpha _i+\frac{A_2}{%
\alpha _i}\right) e^{\alpha _ir}. 
\end{equation}

By the same way the solutions $\psi _{-}^{(2)}$ and $\psi _{-}^{(1){ }}$
could be found

\begin{equation}
\label{18}\psi _{-}^{(2)}=\sum C_i^{\prime }e^{\alpha _ir}, 
\end{equation}

\begin{equation}
\label{19}\psi _{-}^{(1)}=-\frac 1{B_2}\sum C_i^{\prime }\left( \alpha _i+ 
\frac{A_1}{\alpha _i}\right) e^{\alpha _ir}. 
\end{equation}
And the simplest solutions $\psi _{\pm }^{(3)}$ are well known plane waves

\begin{equation}
\label{20}\psi _{\pm }^{(3)}=C_1^{\prime \prime }e^{ip_{\bot }r}+C_2^{\prime
\prime }e^{-ip_{\bot }r}. 
\end{equation}
It is seen from (16) that for $E_2$ spectrum the expression under the root
in $\alpha _{3,4}$ is negative at values $p_{\bot }^2<g^2\left( \tau _1\cos
{}^2\theta +\tau _2\sin {}^2\theta \right) +gH_z^3.$ So, the term $e^{\alpha
_4r}$ in solutions $\psi _{\pm }^{(1),(2)}$ is infinite at $r\rightarrow
\infty .$ This infinity is connected with that the movement of a colored
particle in the field (1) is infinite in space volume, since the energy
spectrum of this particle (11) is continious [9]. So, we shouldn't throw
away this term for normalizibility of the solutions.

In order to find wave functions of states $\psi _1$and $\psi _2$ with
definite energy $E_1$and $E_2$ we have to solve next equations for them 
\begin{equation}
\label{21} 
\begin{array}{c}
\left[ p_1^2+p_2^2+ 
\frac{g^2}4\left( \tau _1+\tau _2\right) I_2+g\sqrt{\tau _1p_1^2+\tau _2p^2+ 
\frac{\left( H_z^3\right) ^2}4}\right] \psi _1=\left( E_1^2-m^2\right) \psi
_1, \\ \left[ p_1^2+p_2^2+\frac{g^2}4\left( \tau _1+\tau _2\right) I_2-g 
\sqrt{\tau _1p_1^2+\tau _2p^2+\frac{\left( H_z^3\right) ^2}4}\right] \psi
_2=\left( E_2^2-m^2\right) \psi _2. 
\end{array}
\end{equation}
Here we have differential operators under the root. Solutions of these
equations are found in appendix and are equal to following expressions%
$$
\psi _{1,2}=-\frac{4\Delta _{1,2}}{a_{1,2}^2\left( c-a_{1,2}\right) }\sin
{}^2\frac{\sqrt{a_{1,2}}}2r\mp \frac{8b\sqrt{-\Delta _{1,2}}}{%
3a_{1,2}^2\left( c-a_{1,2}\right) }\sin {}\frac{\sqrt{a_{1,2}}}2r+\frac{%
4b^2\left( c-a_{1,2}\right) }{9a_{1,2}^2}-\frac{b^2}{4\left(
c-a_{1,2}\right) } 
$$

Thus we can find following superpartner correspondence between states from
same energy branch by use of formulae (7) and solutions (16) -\ (20):

$$
\psi _{+}^{(1)}\rightarrow \Psi _{-}^{(1)}=\sum C_i^{\prime }\left[ \frac{%
ie^{-i\theta }}{B_2}\left( \alpha _i^2+A_1\right) +\frac g2\left( \tau
_1^{\frac 12}-\tau _2^{\frac 12}\right) \right] e^{\alpha _ir} 
$$

$$
\psi _{+}^{(2)}\rightarrow \Psi _{-}^{(2)}=-\sum C_i^{\prime }\left[ \frac
g{2B_2}\left( \tau _1^{\frac 12}+\tau _2^{\frac 12}\right) \left( \alpha _i+ 
\frac{A_1}{\alpha _i}\right) +e^{-i\theta }\alpha _i\right] e^{\alpha _ir} 
$$

\begin{equation}
\label{22}\psi _{+}^{(3)}\rightarrow \Psi _{-}^{(3)}=-ip_{\bot }e^{-i\theta
}\left( C_1^{\prime \prime }e^{ip_{\bot }r}-C_2^{\prime \prime }e^{-ip_{\bot
}r}\right) 
\end{equation}

$$
\psi _{-}^{(1)}\rightarrow \Psi _{+}^{(1)}=\sum C_i\left[ \frac{ie^{i\theta
} }{B_2}\left( \alpha _i^2+A_2\right) +\frac g2\left( \tau _1^{\frac
12}+\tau _2^{\frac 12}\right) \right] e^{\alpha _ir} 
$$

$$
\psi _{-}^{(2)}\rightarrow \Psi _{+}^{(2)}=-\sum C_i\left[ \frac
g{2B_2}\left( \tau _1^{\frac 12}-\tau _2^{\frac 12}\right) \left( \alpha _i+ 
\frac{A_2}{\alpha _i}\right) +e^{i\theta }\alpha _i\right] e^{\alpha _ir} 
$$

$$
\psi _{-}^{(3)}\rightarrow \Psi _{+}^{(3)}=-ip_{\bot }e^{i\theta }\left(
C_1^{\prime \prime }e^{ip_{\bot }r}-C_2^{\prime \prime }e^{-ip_{\bot
}r}\right) . 
$$

At now we can consider particle with $p_3\neq 0.$ Hamiltonian in (5) for
this case is defined by

$$
H^{\prime }=\left( \sigma ^1P_1+\sigma ^2P_2+\sigma ^3p_3\right) ^2 
$$
and keep its spin diagonal form

\begin{equation}
\label{23}H^{\prime }\left( 
\begin{array}{c}
\psi _{+} \\ 
\psi _{-} 
\end{array}
\right) =\left( 
\begin{array}{c}
P_{-}P_{+}+p_3^2\qquad 0 \\ 
0\qquad P_{+}P_{-}+p_3^2 
\end{array}
\right) \left( 
\begin{array}{c}
\psi _{+} \\ 
\psi _{-} 
\end{array}
\right) . 
\end{equation}

This means operators $a_{\pm }$ also keep their anticommutation property and
could be choosn as a fermionic operators. Solutions of (23) equations are
received from (16)-(20) only by multipling on $e^{\pm ip_3z}$. Supercharge
operators defined in the form

\begin{equation}
\label{24} 
\begin{array}{c}
Q_1^{\prime }=Q_1+\sigma ^3p_3 \\ 
Q_2^{\prime }=Q_2+\sigma ^3p_3 
\end{array}
\end{equation}
obey following relations

\begin{equation}
\label{25}\left( Q_1^{\prime }\right) ^2=\left( Q_2^{\prime }\right)
^2=H^{\prime },\ \left[ Q_1^{\prime },H^{\prime }\right] =\left[ Q_2^{\prime
},H^{\prime }\right] =0,\ \left\{ Q_1^{\prime },Q_2^{\prime }\right\}
=2p_3^2\neq 0, 
\end{equation}
last one of which is differ from ordinary supersymmetry algebra. Mutually
hermitian conjugate operators $Q_{\pm }^{\prime }$ constructed in ordinary
way has the form

\begin{equation}
\label{26}Q_{\pm }^{\prime }=\frac 12\left( Q_1^{\prime }\pm iQ_2^{\prime
}\right) =Q_{\pm }+\frac{1\pm i}2\sigma ^3p_3 
\end{equation}
and obey relations

\begin{equation}
\label{27}\ \left\{ Q_{+}^{\prime },Q_{-}^{\prime }\right\} =H^{\prime },\
\left( Q_{\pm }^{\prime }\right) ^2=\pm \frac i2p_3^2, 
\end{equation}
If we define $Q_2^{\prime }=Q_2$, we get ordinary anticommutation $\left\{
Q_1^{\prime },Q_2^{\prime }\right\} =0$ and commutation $\left[ Q_1^{\prime
},H^{\prime }\right] =\left[ Q_2^{\prime },H^{\prime }\right] =0$
supersymmetry relations. But for this definition we have $\left( Q_1^{\prime
}\right) ^2=H^{\prime },\ \left( Q_2^{\prime }\right) ^2=H^{\prime }-p_3^2.$
In addition to this definition, if we define $Q_{\pm }^{\prime }$operators
in following way 
$$
Q_{\pm }^{\prime }=\frac 12\left( Q_1^{\prime }\pm iQ_2^{\prime }\right)
+\frac i2\sigma ^3p_3 
$$
we get for them expressions in (26) with ordinary anticommutation relation
(27). The operators $Q_{\pm }^{\prime }$ contain $\sigma ^3$ matrix. So,
superpartner states in this case are received mixed spin states 
$$
Q_{\pm }^{\prime }\left( 
\begin{array}{c}
\psi _{+} \\ 
\psi _{-} 
\end{array}
\right) =e^{ip_3z}\left( 
\begin{array}{c}
\Psi _{\mp }+ 
\frac{1\pm i}2p_3\psi _{-} \\ -\frac{1\pm i}2p_3\psi _{-} 
\end{array}
\right) +e^{-ip_3z}\left( 
\begin{array}{c}
\Psi _{\mp }- 
\frac{1\pm i}2p_3\psi _{-} \\ \frac{1\pm i}2p_3\psi _{-} 
\end{array}
\right) . 
$$
That means in this three dimensional motion case superpartner states are not
states with definite projection of ordinary spin.

Author thanks to Prof. V.Ch.Zhukovsky for useful remarks.

\section{Appendix}

Let us demonstrate solution of first equation in (21), which in polar
coordinates has got form 
$$
\frac{d^2\psi _1}{dr^2}+b\sqrt{\frac{d^2\psi _1}{dr^2}+c\psi _1}%
+a_1\psi _1=0 
\eqno(A.1)$$
where we have made designations%
$$
a_{1,2}=p_{\bot }^2\pm \sqrt{-b^2p_{\bot }^2+\frac{\left( H_z^3\right) ^2}4}%
,\ b=g\sqrt{-\left( \tau _1\cos {}^2\theta +\tau _2\sin {}^2\theta \right) }%
,\ c=\frac{\left( H_z^3\right) ^2}{4b^2}. 
$$
Equation (A.1) could be written in the form%
$$
\left( \sqrt{\frac{d^2\psi _1}{dr^2}+c\psi _1}+\frac b2\right) ^2-\frac{b^2}%
4+\left( a_1-c\right) \psi _1=0 
$$
square root from which gives 
$$
\sqrt{\frac{d^2\psi _1}{dr^2}+c\psi _1}=\sqrt{\left( c-a_1\right)
\psi _1+\frac{b^2}4}-\frac b2. 
\eqno(A.2)$$
Squaring equation (A.2) again and doing replacement $\psi _1=\frac{\varphi
_1^2}{c-a_1}-\frac{b^2}{4\left( c-a_1\right) }$ we get next equation for $%
\varphi _1$%
$$
2\varphi _1\frac{d^2\varphi _1}{dr^2}+2\left( \frac{d\varphi _1}{%
dr}\right) ^2+a_1\varphi _1^2+b\left( c-a_1\right) \varphi _1-\frac
14b^2\left( 2c-a_1\right) =0. 
\eqno(A.3)$$
Considering $\frac{d\varphi _1}{dr}=f\left( \varphi _1\right) $ and taking
into account $\frac{d^2\varphi _1}{dr^2}=\frac{df}{d\varphi _1}\frac{%
d\varphi _1}{dr}=f^{\prime }f$ in equation (A.3) we get first order
non-linear differential equation for $f\left( \varphi _1\right) $ 
$$
2\varphi _1ff^{\prime }+2f^2=-a_1\varphi _1^2-b\left(
c-a_1\right) \varphi _1+\frac 14b^2\left( 2c-a_1\right) 
\eqno(A.4)$$
Equation (A.4) should be solved without right hand side at first%
$$
2\varphi _1ff^{\prime }+2f^2=0, 
$$
which implies for $f\neq 0$%
$$
\varphi _1f^{\prime }+f=0. 
\eqno(A.5)$$
Equation (A.5) gives $\frac{df}f=-\frac{d\varphi _1}{\varphi _1},$ which
means%
$$
f\cdot \varphi _1=K\left( \varphi _1\right) . 
$$
Taking into account in equation (A.4) last constraint for $f,$ we get
equation for $K\left( \varphi _1\right) $ solution of which is 
$$
K\left( \varphi _1\right) =\sqrt{-\frac{a_1}4\varphi _1^4-\frac{b\left(
c-a_1\right) }3\varphi _1^3+\frac{b^2\left( 2c-a_1\right) }8\varphi _1^2+K_1}%
, 
$$
where $K_1$ is integrate constant. We shall seek solution $\varphi _1$ for $%
K_1=0.$ This means to demand function $f\left( \varphi _1\right) $ to be
finite at some finite $r_n$ points, for which $\varphi _1\left( r_n\right)
=0.$ Using explicit expression of $K\left( \varphi _1\right) $ we find 
$$
\frac{d\varphi _1}{dr}=\sqrt{-\frac{a_1}4\varphi _1^2-\frac{%
b\left( c-a_1\right) }3\varphi _1+\frac{b^2\left( 2c-a_1\right) }8}. 
\eqno(A.6)$$
Integration of (A.6) is carry out for $\Delta _1<0$ and $\Delta _1>0$ values
of $\Delta _1=-\frac{a_1b^2}8\left( 2c-a_1\right) -\frac{b^2}9\left(
c-a_1\right) $ and gives same result for $\varphi _1$%
$$
\varphi _1=\frac{2\sqrt{-\Delta _1}}{a_1}\sin \frac{\sqrt{a_1}}2r-\frac{%
2b\left( c-a_1\right) }{3a_1}+Const. 
$$
Only solution with $Const.=0$ obey initial equation. Same calculation for
second equation of (21) gives%
$$
\varphi _2=\frac{2\sqrt{-\Delta _2}}{a_2}\sin \frac{\sqrt{a_2}}2r+\frac{%
2b\left( c-a_2\right) }{3a_2}. 
$$
From these expressions we can find $r_n$ points of $\varphi _{1,2}$ 
$$
r_n=\frac 2{\sqrt{a_{1,2}}}\left( \arcsin \pm \frac{c-a_1}{3\sqrt{-\Delta
_{1,2}}}+2\pi n\right) . 
$$

\newpage\

\end{document}